\documentclass{aa} % if the references are not structured 
\usepackage{graphicx}
\usepackage{txfonts}
\usepackage{hyperref}
\hypersetup{%
    colorlinks=true, 
    citecolor=blue,
    filecolor=blue,
    linkcolor=blue,
    urlcolor=black
    } 

\begin{document} 

   \title{GRB\,080928 afterglow imaging and spectro-polarimetry}

   \author{R. Brivio
          \inst{1,2}\fnmsep\thanks{email: riccardo.brivio@inaf.it}
          \and
          S. Covino\inst{2}
          \and
          P. D'Avanzo\inst{2}
          \and
          K. Wiersema\inst{3}
          \and
          J. R. Maund\inst{4}
          \and
          M.G. Bernardini\inst{2}
          \and
          S. Campana\inst{2}
          \and\\
          A. Melandri\inst{2}
          }

   \institute{Università degli Studi dell’Insubria, Dipartimento di Scienza e Alta Tecnologia, via Valleggio 11, 22100 Como, Italy 
         \and
             INAF–Osservatorio Astronomico di Brera, via E. Bianchi 46, 23807 Merate (LC), Italy
            % \email{stefano.covino@inaf.it, paolo.davanzo@inaf.it}
        \and
            Physics Department, Lancaster University, Lancaster, LA1 4YB, UK
        \and
            Department of Physics and Astronomy, University of Sheffield, Hicks Building, Hounsfield Road, Sheffield S3 7RH, UK
             }

   \date{Received xxx; accepted yyy}

% \abstract{}{}{}{}{} 
% 5 {} token are mandatory
 
  \abstract
  % context heading (optional)
  % {} leave it empty if necessary  
   {Among the large variety of astrophysical sources that we can observe, gamma-ray bursts (GRBs) are the most energetic of the whole Universe. Their emission peaks in the $\gamma$-ray band, with a duration from a fraction of a second to a few hundred seconds, and is followed by an afterglow covering the whole electromagnetic spectrum. The definition of a general picture describing the physics behind GRBs has always been a compelling task, but the results obtained so far from observations have revealed a puzzling landscape. The lack of a clear, unique paradigm calls for further observations and additional, independent techniques for this purpose. Polarimetry constitutes a very useful example as it allows us to investigate some features of the source such as the geometry of the emitting region and the magnetic field configuration.}
  % aims heading (mandatory)
   {To date, only a handful of bursts detected by space telescopes have been accompanied by ground-based spectro-polarimetric follow-up, and therefore such an analysis of more GRBs is of crucial importance in order to increase the sample of bursts with multi-epoch polarisation analysis. In this work, we present the analysis of the GRB\,080928 optical afterglow, with observations performed with the ESO-VLT FORS1 instrument.}
  % methods heading (mandatory)
   {Starting from raw data taken in the imaging polarimetry (IPOL) and spectro-polarimetry (PMOS) modes, we performed data reduction, followed by the photometric analysis of IPOL data, taken $\sim$ 14 and $\sim$ 40 hours after the burst detection, and spectroscopy of PMOS data ($t$ $\sim$ 14.95 h). After computing the reduced Stokes parameters $Q/I$ and $U/I$, which describe the linear polarisation of the emitted radiation, we obtained the polarisation degree for the three observing epochs.}
  % results heading (mandatory)
   {We find that the GRB optical afterglow was not significantly polarised on the first observing night. The polarisation degree ($P$) grew on the following night to a level of $P\sim4.5$\%, giving evidence of polarised radiation at a 4$\sigma$ confidence level. The GRB\,080928 light curve is not fully consistent with standard afterglow models, making any comparison with polarimetric models partly inconclusive. The most conservative interpretation is that the GRB emission was characterised by a homogeneous jet and was observed at an angle of $0.6<\theta_{obs}/\theta_{jet}<0.8$. Moreover, the non-zero polarisation degree on the second night suggests the presence of a dominant locally ordered magnetic field in the emitting region.}
  % conclusions heading (optional), leave it empty if necessary 
   {}


   \maketitle
%
%-------------------------------------------------------------------
\section{Introduction}\label{sec:intro}

    Gamma-ray bursts (GRBs) represent the brightest events we can observe in the Universe, reaching 10$^{54}$ erg/s of isotropic equivalent energy. They are made up of a bright, prompt phase peaking in the $\gamma$-ray band, and a long-lasting (from hours to days or even weeks and months), fading afterglow that covers the electromagnetic spectrum at all wavelengths \citep{Meszaros&Rees97, Sarietal98, Piran99}. The observation of the prompt phase duration of several bursts revealed a bimodal distribution (\citealt{Kouveliotouetal93}) that led to a classification into two categories: long (LGRBs) and short (SGRBs), with the separation at about 2 s. These phenomena are the result of the collapse of a massive star or the merger of compact binaries, with a significant fraction of the long GRBs associated with core-collapse supernovae \citep[see][for a review]{Canoetal17}. Accretion onto the resulting compact object (either a black hole or a neutron star) produces powerful ultra-relativistic jets generating the prompt emission through dissipation processes like shocks or magnetic reconnection \citep[for a review of GRB physics, see e.g.][]{Zhang&Meszaros04, Gehrels09, Kumar&Zhang15}. The afterglow is instead produced when the resulting rapidly expanding ejecta of a GRB collide with the surrounding medium. As the collision-driven afterglow emerges, shocks are formed: one forward-propagating into the external medium and another shorter-lived reverse shock propagating backward into the jet \citep{Sari&Piran99_RS, Kobayashi_RS}. The relative importance of these shocks is set by some micro-physical parameters, depending on the magnetic field and electron energies.
    
    At present, a large number of GRBs have been observed, especially after the launch of dedicated missions like the Neil Gehrels Swift Observatory \citep[hereafter \textit{Swift},][]{SWIFT} in 2004 and Fermi Gamma-Ray Space Telescope, with its Large Area Telescope \citep[LAT, ][]{FERMI-LAT} and Gamma-ray Burst Monitor \citep[GBM, ][]{FERMI-GBM} in 2008. Despite the effort in this field, a unique and general model  reconciling all observations and inferred features of detected bursts has proven elusive. Polarimetry can allow us to prove that the fireball is beamed, to constrain the orientation of the jet with respect to the line of sight, and possibly to determine the jet geometry as well as the configuration of the magnetic fields dominating in the emitting region \citep[for a review, see][]{Covino&Gotz16}. Some degree of polarisation is expected to emerge in the optical flux of GRBs as a signature of synchrotron radiation \citep{Meszaros&Rees97}, and the first successful polarisation measurement was achieved for the optical afterglow (OA) of GRB\,990510 \citep{Covinoetal99,Wijersetal99}. As a general rule, some degree of asymmetry in the expanding fireball is necessary to produce some degree of polarised flux. \citet{Gruzinov&Waxman99} argued that if the magnetic field is globally random but with a large number of patches where the magnetic field is instead coherent, a polarisation degree of up to $\sim$10\% is expected, especially at early times. High levels of polarisation in the early afterglow were observed for some GRBs, such as GRB\,090102, GRB\,091208, and GRB\,120308A \citep[see respectively,][]{Steeleetal09, Ueharaetal12, Mundelletal2013}. \cite{GL99} and, independently, \cite{Sari99}, considered a geometrical setup in which a beamed fireball is observed slightly off-axis. This break of symmetry again results in a significant polarisation degree. Their model also predicts a testable variation of the polarisation degree and position angle associated with the evolution of the afterglow light curve.
    
    A relatively poorly explored polarimetric probe of afterglow physics is spectro-polarimetry, which extends the available polarisation measurements to a wide range of wavelengths. Multi-wavelength, simultaneous detection of polarised flux offers an additional and efficient way to explore the afterglow physics both at early and late times and from both the forward and the reverse shock. Spectro-polarimetry adds some diagnostic power especially at optical wavelengths, in particular if any of the synchrotron break frequencies (e.g. the synchrotron cooling frequency) lie within the optical band.
    Spectro-polarimetry is also crucial to quantify the polarisation induced by aligned dust along the line of sight in the GRB host galaxy and in our own galaxy. Indeed, it is now well-established that the optical afterglow radiation is mainly produced via synchrotron emission, whose associated polarisation is expected to be wavelength independent, while dust-induced polarisation makes the afterglow polarisation $\lambda$-dependent. Therefore, spectro-polarimetry is the best technique to quantify this contribution, which is likely to play a non-negligible role in the polarisation distribution of afterglows and their physical interpretation. To date, only a few afterglows have been observed with spectro-polarimetry; for example GRB\,020813 \citep{Barthetal03}, GRB\,021004 \citep{Wangetal03}, GRB\,030329 \citep{Greineretal04}, and GRB\,191221B \citep{Buckleyetal21}. 
    
    A crucial parameter for a more complete physical interpretation of the event is the jet break time, that is, the time at which we observe an achromatic break in the afterglow light curve. Despite not being trivial, in some cases a polarisation detection before and after the jet break was achieved, leading to a more precise modelisation of the afterglow: a decreasing polarisation from $P =  2.26\%$ to $P = 1.18\%$ was identified in the GRB\,020813 optical afterglow \citep{Barthetal03, Gorosabel020813, Lazzati020813}, while an upper limit was given from a detection almost coincident with the jet break for GRB\,071010A \citep{Covino071010A}; polarisation measurements before and after the jet break were also achieved for GRB\,091018 \citep{Wiersemaetal12} and GRB\,121024A \citep{Wiersemaetal14}.
    
    In this work, we analyse imaging polarimetry and spectro-polarimetry data of the optical afterglow of GRB\,080928, an event that has not yet been properly analysed and published\footnote[2]{in \cite{Covino&Gotz16} results from a very preliminary analysis are reported.}. In Sect.\hyperref[sec:GRB]{2} we report information about GRB\,080928, and in Sect. \hyperref[sec:obs]{3} the observations and data analysis procedure are described. In Sect. \hyperref[sec:res]{4} a full discussion is presented, while our main conclusions are summarised in Sect. \hyperref[sec:disc]{5}.
   
%--------------------------------------------------------------------
\section{GRB\,080928}\label{sec:GRB}

    GRB\,080928 was first detected by the Burst Alert Telescope \citep[BAT, ][]{BAT} on board $\textit{Swift}$ at $t_0$ = 15:01:32.86 UT on 2008 September 28 \citep{SWIFT_GCN}. Its prompt phase lasted 112 s, making it a candidate for the Long GRB class. The main burst emission also triggered the Gamma-Ray Burst Monitor on board Fermi \citep{FERMI_GCN}, while the INTEGRAL satellite was passing through the South Atlantic Anomaly (SAA) during the time of GRB\,080928 and therefore could not observe the burst. $\textit{Swift}$ slew to point towards the emitting region with its X-Ray Telescope \citep[XRT, ][]{XRT} 170 seconds after the BAT trigger, observing until 2.7 days after the burst detection. The optical counterpart was identified by the UV/Optical Telescope \citep[UVOT, ][]{UVOT} from 3 min after the trigger and was located at the coordinates (J2000) R.A. = 6h 20m 16.83s, Dec = -55$^{\circ}$11'58."9, with an uncertainty of 0.5", and its redshift was found to be $z = 1.6919$ \citep{REDSHIFT}. The prompt emission was characterised by a precursor started at $t_0-90$ s and some peaks of variable intensity: the main GRB emission started at $t_0+170$ s, with two peaks at 204 and 215 s. $\textit{Swift}$ also detected a third, fainter peak at 310 s, and stopped the observations at $t_0+400$ s. The Fermi/GBM light curve shows only a single pulse corresponding to the main peak detected by $\textit{Swift}$ ($t_{0,\text{GBM}} = t_0+204$ s). The optical emission was studied in detail by \citet{ROSSI_A11}, where the X-rays and optical light curves were published.
    The fluence in the prompt phase in 15-150 keV band (that of $\textit{Swift}$/BAT) is (2.1 $\pm$ 0.1)$\times$10$^{-6}$ erg/cm$^2$, while the 1-sec peak photon flux, measured in the same band, is 2.1 $\pm$ 0.1 photons/cm$^2$/sec (errors at 90\% confidence level).
    
    Ground-based follow-up observations were performed by the ROTSE-IIIa 0.45m telescope in Australia \citep{ROTSE_GCN}, the Watcher telescope in South Africa \citep{WATCHER_GCN}, the MPG/ESO 2.2m telescope on La Silla, Chile, equipped with the multichannel imager GROND \citep{GROND_GCN}, and ESO-VLT at the Paranal Observatory, Chile \citep{FORS2_GCN}.

\section{Observations and data analysis}\label{sec:obs}

    %--------------------------------------------------- OA mag from acquisition image
    \begin{table}
        \caption{ESO-VLT FORS1 optical afterglow magnitudes}
        \centering
        \begin{tabular}{ccc}
        \hline
        \noalign{\smallskip}
        $t - t_0$ [h]& V magnitude& $\sigma_{V_{mag}}$\\
        \hline
        \hline
        \noalign{\smallskip}
        13.93& 20.89& 0.03\\
        13.97& 20.85& 0.03\\
        14.85& 21.07& 0.09\\
        15.73& 21.03& 0.09\\
        39.30& 22.71& 0.11\\
        40.30& 22.77& 0.13\\
        41.20& 22.65& 0.11\\
        \noalign{\smallskip}
        \hline
        \noalign{\smallskip}
        \end{tabular}
        \tablefoot{V-band magnitudes for the optical afterglow computed from the acquisition images taken before polarimetry epochs and calibrated using APASS standard stars, as described in the text.}
        \label{acq_im_lc}
    \end{table}

    %------------------------------------------- Magnitudes
    \begin{figure*}
    \centering
    \includegraphics[width=15cm]{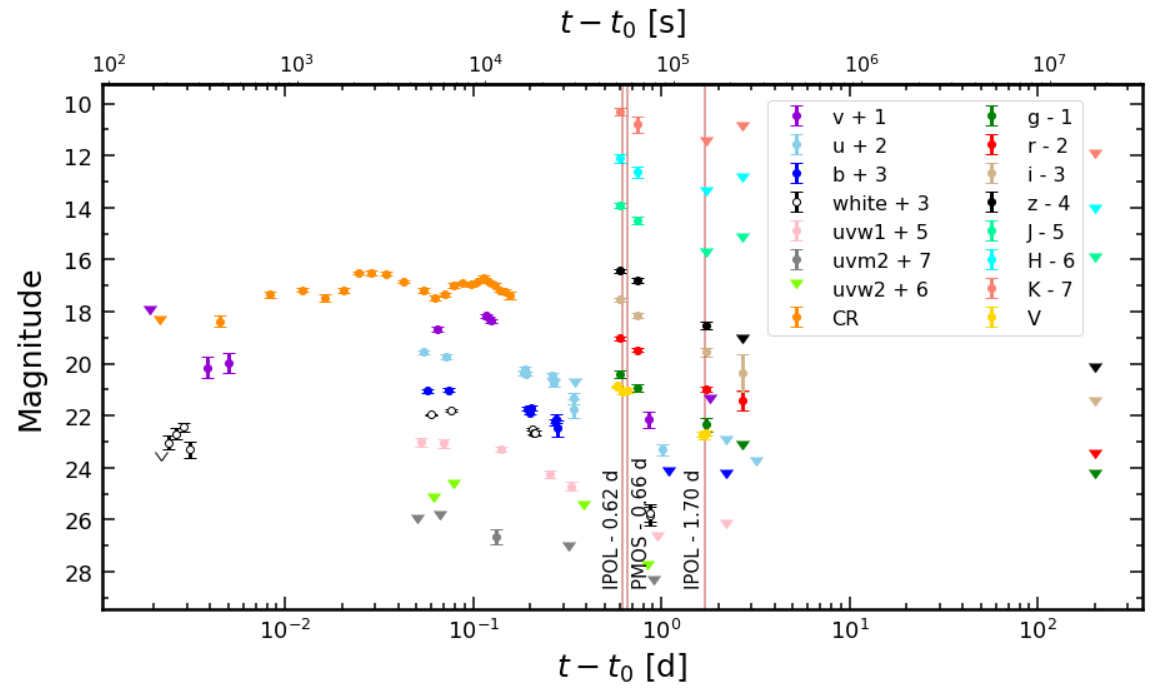}
    \caption{Complete photometric optical/NIR data set of the afterglow of GRB\,080928 from $ \textit{Swift}$/UVOT (all but the last filter in the first column of the legend), ROTSE-IIIa (CR, i.e. unfiltered R-equivalent Vega magnitudes), GROND (all but the last filter in the second column of the legend), and VLT-FORS (V filter) observations. All magnitudes are given in the Vega system, with colours shifted by the values given in the legend for clarity. Downward pointing triangles indicate upper limits; \textit{uvw2} was the only filter in which only upper limits could be derived. The vertical lines mark the time of our polarimetric measurements.}
    \label{mag}
    \end{figure*}
    
     \begin{figure}
    \centering
    \includegraphics[width=9cm]{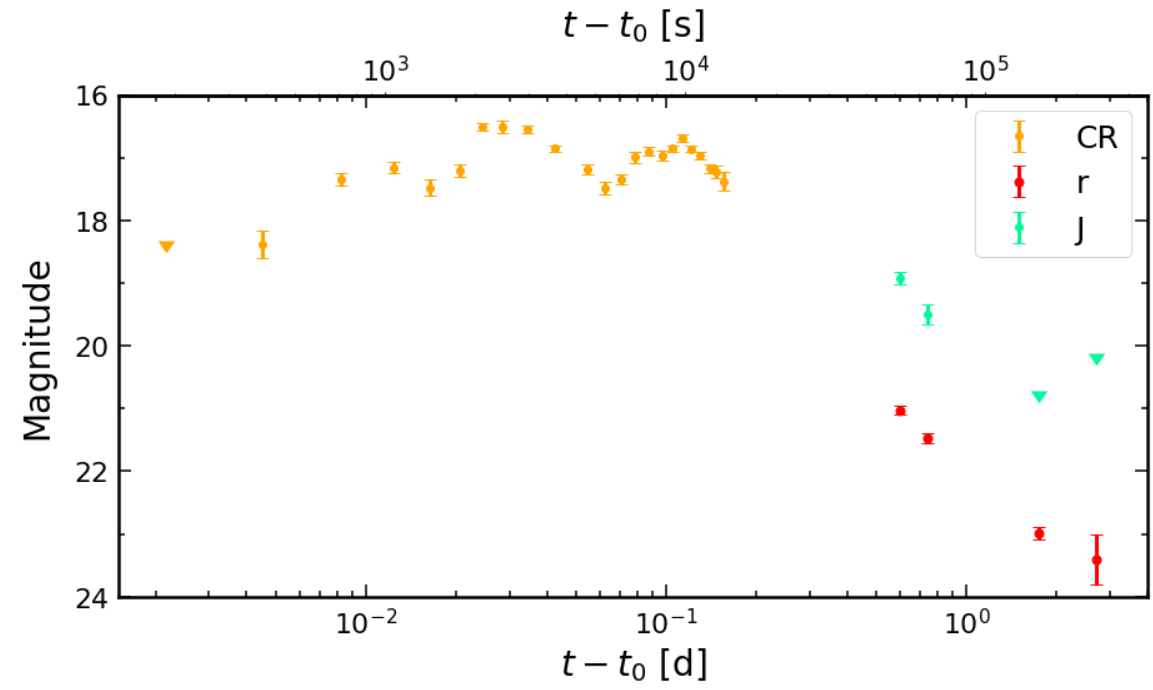}
    \caption{R-band and J-band points from the complete data set shown in Fig.\ref{mag}, without shift in magnitude, to show the temporal evolution of the optical-NIR afterglow of GRB\,080928. Optical r-band flux is nearly constant up to $t-t_0 \sim$ 10$^{-1}$ d, before it decreases as expected for standard afterglows; NIR J-band observations show a similar decrease even if two points and only two upper limits were obtained.}
    \label{mag_bis}
    \end{figure}
%-----------------------------------------------------------------
\renewcommand{\thefootnote}{\arabic{footnote}}
    
    Observations of GRB\,080928 were obtained by ESO VLT–UT2 (Kueyen), which is equipped with the Focal Reducer/low dispersion Spectrometer (FORS1) with the V\_HIGH filter\footnote{ESO filter number +114, for transmission curves see \url{https://www.eso.org/sci/facilities/paranal/instruments/fors/inst/Filters/curves.html}} ($\lambda_0$=5610 \AA, FWHM=1230 \AA) in the imaging polarimetry mode; the 300V grism (with order sorting filter GG375) and a 1.5" slit width were adopted in the spectro-polarimetry mode, with a spectrum covering the range from 3800 $\AA$ to 7600 $\AA$. Observations were all obtained with the E2V blue-optimised CCD mounted on the instrument, and the 2x2 binning readout mode of the CCD was adopted. All raw data, including both target observations and calibration data, were downloaded from the ESO raw data archive\footnote{\url{http://archive.eso.org/cms.html}}. The burst is of particular interest because both optical and X-ray emissions were detected when the GRB was still radiating in the gamma-ray band. This makes it one of the rare cases \citep[e.g. GRBs 041219A, 050820A, 051111, 061121;][]{Shen&Zhang09} where a broad-band spectral energy distribution (SED) from about 1 eV to 150 keV can be constructed for the prompt emission phase.
    
    Our dataset includes two observing runs with the imaging polarimetry (IPOL) mode divided into two nights: these started $\sim$ 14 hours after the GRB trigger on 2008 September 28 night, and the same setup was repeated three times on the following night from $t - t_0 \sim$ 40 h. The whole IPOL dataset is reported in Table \ref{data_ipol}. In addition, one observation in the spectro-polarimetry (PMOS) mode was performed from $t - t_0 \sim$ 14.95 h (see Table \ref{seeing}). All the observing blocks reported in Table \ref{data_ipol} are made up of two subsets, each containing four exposures at four different angles (0$^{\circ}$, 22.5$^{\circ}$, 45$^{\circ}$, 67.5$^{\circ}$) of the half-wave plate in the instrumental setting of FORS1. Imaging polarimetry is achieved by the use of a Wollaston prism splitting the image of each object in the field into the two orthogonal polarisation components appearing in adjacent areas of the image. A strip mask is used in the focal area of the instrument to avoid overlapping of the two beams of polarised light on the CCD. In this way, for each position angle $\phi$/2 of the half-wave plate rotator, we obtain two simultaneous images of cross-polarisation at angles $\phi$ and $\phi$+90$^{\circ}$. The dataset also includes some standard stars to be analysed: two polarised, Vela1\,95 (observed in both modes) and NGC-2024 (in IPOL mode), in order to fix the offset between the polarisation and the instrumental angles, and one unpolarised, WD-2149+021 (in PMOS mode), to check for possible spurious instrumental contributions to the total polarisation degree.
    
    Data analyses were performed for all observations by means of specific software depending on the data type. The reduction was carried out with the ESO-Eclipse package \citep[version 5.0.0,][]{ECLIPSE} for IPOL images and with the ESO-Reflex pipeline in PMOS mode \citep[version 2.11.3,][]{REFLEX} for spectro-polarimetric data. For both of them, after bias subtraction, non-uniformities were corrected using flat-fields obtained without the Wollaston prism \citep[see e.g.][]{Patat&Romaniello06}. 
    
    The flux of each point source in the field of view of IPOL data was derived by means of aperture photometry by the Graphical Astronomy and Image Analysis (GAIA) tools \citep{GAIA}, with apertures chosen so as to be at least a few times the seeing. The background was estimated with annuli of radii from two to three times the seeing and applying a sigma clipping of the counts computed inside. Each pair of simultaneous measurements at orthogonal angles was used to compute the $Q, U$ Stokes parameters. This technique removes any difference between the two optical paths (ordinary and extraordinary ray). Moreover, being based on relative photometry in simultaneous images, our measurements are insensitive to intrinsic variations in the optical transient flux. In addition to the OA, some bright, nearby field stars were studied in order to look for possible spurious contribution, because stars are typically unpolarised sources, except for some polarisation induced by dust either in the host galaxy or in the Milky Way. All the field stars we studied are less than 1$'$ from the position of the optical afterglow, meaning that the instrumental polarisation depending on the distance from the optical axis of the instrument can be neglected \citep{Patat&Romaniello06}.
    
    For the spectropolarimetric data, 1D spectra were extracted from reduced 2D images through ESO-Midas\footnote{\url{https://www.eso.org/sci/software/esomidas/}} (version 19FEB) after checking the width of the signal by fitting the spatial profile with a Gaussian distribution. This allowed us to also compute the corresponding seeing; the values are reported in Table \ref{seeing}. We then rebinned the extracted spectra so as to obtain a larger signal-to-noise ratio (S/N) in each bin: we chose a total of 50 bins from 3801.65 to 7596.65 \AA, giving a spectral dispersion of 75.96 \AA/bin.
    
    The reduced Stokes parameters $Q/I$ and $U/I$ describing the linear polarisation of the radiation were derived using the following formulae:
    
    \begin{eqnarray}
    \begin{aligned}
    \frac{Q}{I} &= \frac{1}{2}\bigg(\frac{f_o-f_e}{f_o+f_e}\bigg|_{0^{\circ}}-\frac{f_o-f_e}{f_o+f_e}\bigg|_{45^{\circ}}\bigg)\\
    \frac{U}{I} &= \frac{1}{2}\bigg(\frac{f_o-f_e}{f_o+f_e}\bigg|_{22.5^{\circ}}-\frac{f_o-f_e}{f_o+f_e}\bigg|_{67.5^{\circ}}\bigg)
    \end{aligned}
    \label{qu_reduced}
    ,\end{eqnarray}
    \\
    which make use of all data obtained at the four angles of the retarder plate. The subscripts $_o$ and $_e$ represent the ordinary and extraordinary rays in which the incoming radiation is split, respectively. We computed the reduced Stokes parameters from the average of the results obtained for all observing runs on the same night, and so obtained two final values for imaging polarimetry data, one per night, and one result in the spectro-polarimetry mode. Errors on the reduced Stokes parameters are computed via standard propagation theory.
    
    In addition, we computed the magnitude of the OA corresponding to the polarimetry epochs: before each observing run, acquisition images of the sky were obtained, from which we could derive magnitudes in V band. The data were calibrated using field stars from the AAVSO Photometric All-Sky Survey Data Release 9 \citep[APASS DR9, ][]{APASS}: we made use of some field stars, whose FORS V-band magnitudes were compared with V magnitudes from APASS. In this way, we found the appropriate correction to be applied to the afterglow magnitude by fine-tuning the zero-point on each acquisition image. The results obtained for the optical transient are shown in Table \ref{acq_im_lc}, and were added to the whole data set of optical and near-infrared observations for GRB\,080928 afterglow, including the ground-based follow-up and the $\textit{Swift}$/UVOT observations. Data were taken from Tables A.1, A.2, and A.3 in \citet{ROSSI_A11}. The complete set is plotted in Fig.\ref{mag}; moreover, for the sake of clarity regarding the afterglow time evolution, we show R-band and J-band data only in Fig.\ref{mag_bis}: after an initial, almost constant behaviour, the flux undergoes a power law-like decrease from $t - t_0 \sim$10$^{-1}$ d.

\section{Results}\label{sec:res}

    %
%---------------------------------------- Q,U PLOT - NIGHT 1 
    \begin{figure}
    \centering
    \includegraphics[width=8cm]{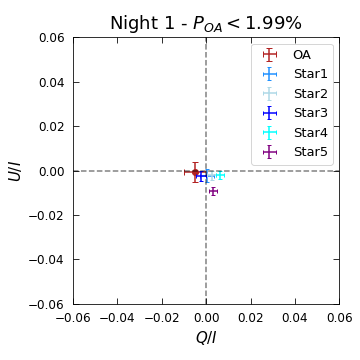}
    \caption{$Q/I, U/I$ plot from the IPOL observation on the night of 2008 September 28. The position of the optical afterglow (red bars) does not differ markedly from the other points, which represent some bright field stars observed near the transient and expected to be unpolarised. Therefore, the polarisation of the OA is consistent with zero, and only a 3$\sigma$ upper limit is given (see value above the plot).}
    \label{QU1}
   \end{figure}
  
%-------------------------------------- Q,U PLOT - NIGHT 2 
    \begin{figure}
    \centering
    \includegraphics[width=8cm]{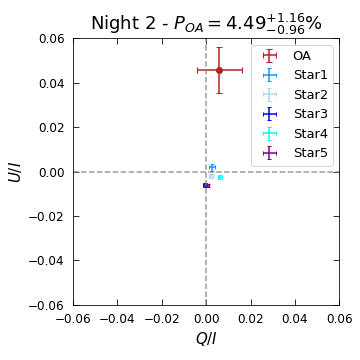}
    \caption{$Q/I, U/I$ plot from IPOL observation on the night of 2008 September 29. The position of the optical afterglow (red bars) is now far away from the cluster formed by the field stars, which are still expected to be unpolarised. For this reason, at a confidence level slightly larger than 4$\sigma$ (see values above the plot), we can state that the optical transient is polarised.}
   \label{QU2}
   \end{figure}

%-------------------------------------- Q,U SPECTRA - NIGHT 1    
   \begin{figure}
    \centering
    \includegraphics[width=9cm]{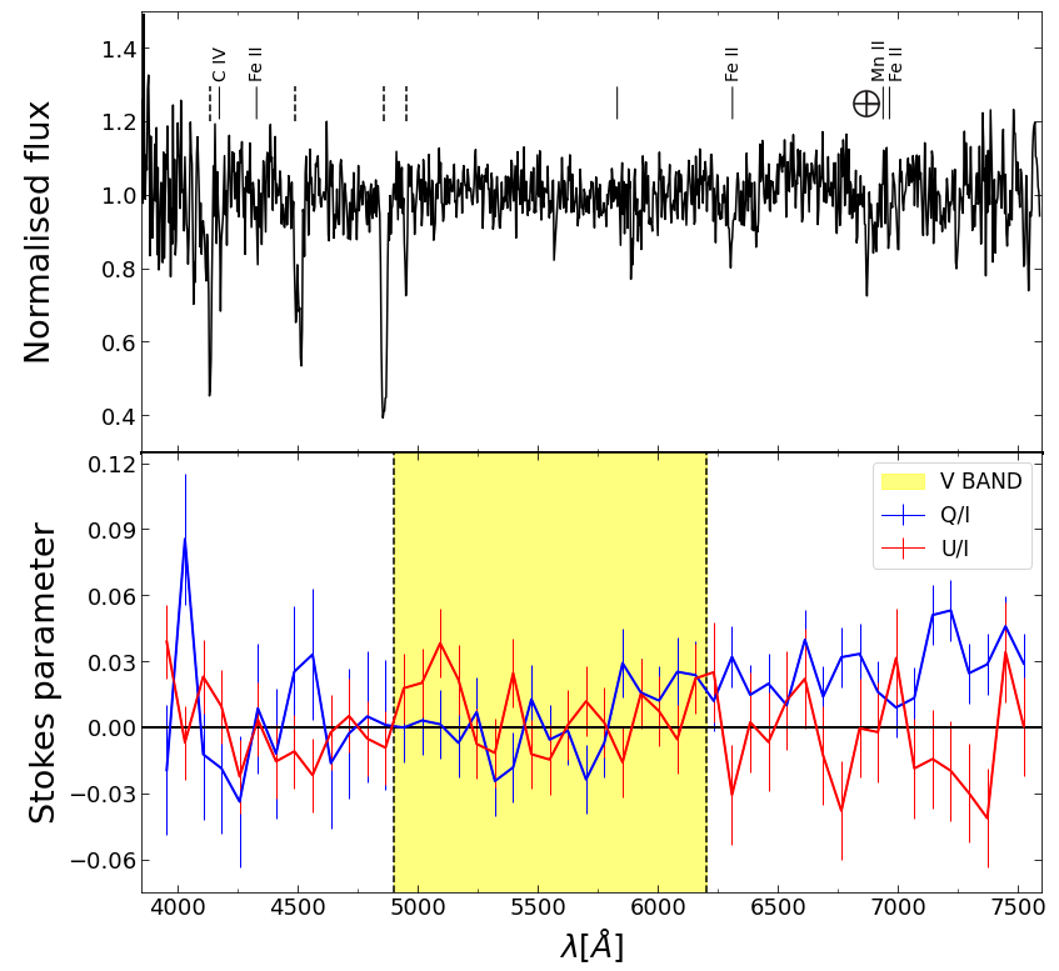}
        \caption{Results of GRB 080928 spectroscopic analysis.\\
        \textit{Top}.GRB\,080928 normalised spectrum obtained from FORS1/PMOS data taken from $t \sim$ 14.95 h after the burst on 2008 September 28. Spectral lines from GRB\,080928 absorption system at $z$=1.6919 are marked with vertical solid lines, while the ones from an intervening system at $z$=0.7359 are represented by vertical dashed lines \citep[for details, see][]{REDSHIFT}.\\
        \textit{Bottom}. $Q/I$, $U/I$ Stokes parameters spectra derived from FORS1/PMOS data. The highlighted yellow region corresponds to the V band in which IPOL data were taken in order to perform a consistent comparison between data taken with the two observing modes. The spectra have been rebinned as described in the text.}
    \label{qu_spec}
    \end{figure}

%----------------------------------------- RESULTS
    \begin{table*}
    \caption{Results of GRB 080928 optical afterglow polarimetric observations}
    \centering
    \begin{tabular}{cccccc}
    \hline
    \noalign{\smallskip}
    $t - t_0$ [d]& MODE& $Q/I$& $U/I$& $P$ [\%]& $\theta$ [$^{\circ}$]\\
    \noalign{\smallskip}
    \hline
    \hline
    \noalign{\smallskip}
    0.62& IPOL& $-$0.0049 $\pm$ 0.0050& $-$0.0008 $\pm$ 0.0045& < 1.99& $-$\\
    \noalign{\smallskip}
    0.66& PMOS& 0.0031 $\pm$ 0.0034& 0.0037 $\pm$ 0.0038& < 1.57& $-$\\
    \noalign{\smallskip}
    1.70& IPOL& 0.0059 $\pm$ 0.0101& 0.0457 $\pm$ 0.0104& 4.49$^{+1.16}_{-0.96}$& 41.3 $\pm$ 6.3\\
    \noalign{\smallskip}
    \hline
    \noalign{\smallskip}
    \end{tabular}
    \tablefoot{Summary of the results in both imaging and spectro-polarimetry mode. The latter were obtained by integrating the spectrum over the V band. Errors on the reduced Stokes parameters and the position angle were computed via propagation theory; those on $P$ were obtained after bias correction when appropriate (see text). Uncertainties are at 1$\sigma$, while upper limits are at 3$\sigma$.}
    \label{result}
    \end{table*}
    
%------------------------------------------- FINAL PLOT: P, theta vs. time
    \begin{figure*}
    \centering
    {\includegraphics[width=11cm]{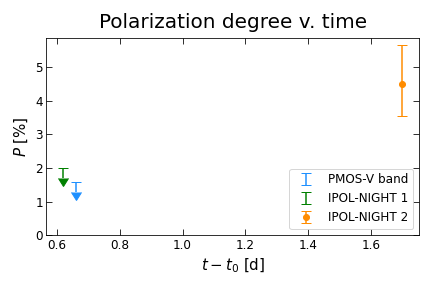}}
    \caption{Evolution of the polarisation degree $P$ of GRB\,080928, both in imaging polarimetry and spectropolarimetry mode. The afterglow is unpolarised at early times ($\sim$ 14-15 hours after the trigger), while $P$ significantly increases about one day after (at $t-t_0 \sim$ 40.7 h). As we could only derive upper limits on the first night, we cannot properly analyse the evolution of the position angle with time.
            }
    \label{p_t}
    \end{figure*}

%--------------------------------------------------------------

    After computing and averaging the Stokes parameters, we obtained (for IPOL data) the $Q/I, U/I$ plots reported in Fig.\ref{QU1} and Fig.\ref{QU2}. They include the OA and the field stars, with bars representing 1$\sigma$ errors. All the field stars are isolated, unsaturated in every epoch and at least comparable in brightness to the afterglow in the first epoch. During the first night, we notice that the location of the optical afterglow does not differ significantly from the locations of the fields stars, and therefore we do not have evidence of polarised radiation emitted $\sim$14 hours after the burst. On the second night, instead, the OA is far from the cluster of field stars on the $Q/I, U/I$ plot, providing possible evidence of polarised radiation. As for the spectra, we summed all results obtained in PMOS mode and obtained the spectra reported in Fig.\ref{qu_spec}. To compute the polarisation degree, we considered only the range in wavelength corresponding to the one of the V\_HIGH filter used for IPOL observations, so as to make a comparison. 
    
    From the analysis of the standard stars observed in the imaging polarimetry mode, we derived a polarisation angle for Vela1\,95 of $\theta_{\text{Vela}} = 172.33 \pm 0.27$, while the expected one is $\theta_{\text{exp,Vela}} = 172.1 \pm 0.1$\footnote{\url{https://www.eso.org/sci/facilities/paranal/instruments/fors/inst/pola.html}}; by applying the same analysis to NGC-2024 we found $\theta_{\text{NGC}} = 136.55 \pm 0.26$, with an expected value of $\theta_{\text{exp,NGC}} = 135.94 \pm 0.2$ \citep{STD_POL}. Therefore, the analysis of polarimetric standard stars revealed consistency with expected values within 1$\sigma$ for Vela1 95 and 1.5$\sigma$ for NGC 2024. Moreover, the average polarisation of the field stars is consistent with zero: on the first night, we obtained <$Q/I$> = 0.0025 $\pm$ 0.0027, <$U/I$> = 0.0016 $\pm$ 0.0012, while on the second night we derived <$Q/I$> = 0.0017 $\pm$ 0.0016, <$U/I$> = 0.0011 $\pm$ 0.0004. The degree $P$ and angle $\theta$ of polarisation are obtained from the measurements of $Q/I$ and $U/I$ for the OA [$P=\sqrt{Q^2+U^2}/I$, $\theta=\frac{1}{2}\arctan(U/Q)$] after correcting for the polarisation induced by the instrument or by the local interstellar matter (<$Q/I$>, <$U/I$>). Moreover, for any low level of polarisation, ($P/\sigma_P \leq$ 4), the distribution function of $P$ and of $\theta$ is no longer normal and that of $P$ becomes skewed. A correction taking into account this bias is required, and we adopted the modified asymptotic estimator defined by \cite{Plaszczynskietal14} to derive the correct value of the polarisation degree $P$. In this way, we obtained the final results for the polarisation of the OA and report them in Table \ref{result}. As we found very low values of $P/\sigma_P$ for the observations taken on the first night, we report only 3$\sigma$ upper limits for the polarisation degree $P,$ while an evaluation of the position angle evolution was not possible. We notice that there is no significant polarisation on the first night, as confirmed also from PMOS results, while we found polarisation on the following night at more than 4$\sigma$ confidence level (CL). We also combined the results obtained on the first night in both IPOL and PMOS mode to obtain an average value for the polarisation detected on 2008 September 28, yielding a 3$\sigma$ upper limit of $P<1.06\%$. The evolution of the polarisation with time is shown in Fig.\ref{p_t}. The polarisation is consistent with zero on the first night, as already pointed out, and it increases towards the value obtained for the second night. No further observations were available to follow the later evolution of the polarisation curve.
    
    As already pointed out in Sect.\ref{sec:intro}, the dust present in both the GRB host galaxy and the Milky Way (MW) can contribute to the total polarisation degree. Starting from the host extinction estimated by \cite{ROSSI_A11}, we tried to put some limits on the polarisation contribution from the dust in the host galaxy: different dust recipes were considered (SMC, LMC, and MW for Small and Large Magellanic Clouds and Milky Way extinction laws, respectively; \citealt{Gordonetal03_extinction, Gordonetal09_extinction}) in order to try to derive the local extinction $A_V$, which was found to be in the 0.04-0.37 mag range, depending on the specific model adopted. It is possible to put an upper limit on the contribution of the galactic interstellar polarisation: $P_{\text{ISP}}(\%) \leq 9.0\times E(B-V)$. If we assume a MW-like behaviour of the dust in the host galaxy, we can derive $E(B-V) = A_V/R_V \leq 0.12$, where we have considered the larger estimate for $A_V$ (0.37) and $R_V = 3.1$, that is, the average value for our galaxy. This yields a maximum contribution from the GRB host galaxy dust of $P_{\text{HG}}\sim$ 1.1\%, which is significantly lower than the upper limit we derived. In order to investigate the significance of this possible contribution, we tried a fit of the polarisation spectrum in a Bayesian framework with either a Serkoswki law \citep{Serkowskietal75} or the predictions in the optical band for a standard afterglow, that is, a constant value. A fit with a Serkowski law is satisfactory and gives  $\lambda_{\rm max}\sim 0.8\,\mu$m, likely driven by the apparent polarisation increase in the reddest part of the spectrum (Fig.\,\ref{qu_spec}). A large value for $\lambda_{\rm max}$ is not unprecedented in young stellar population environments \citep[e.g.][]{Patatetal15}. However, a constant (afterglow) value for the polarisation also provides a good fit and the preference for the more complex model with respect to the simpler afterglow, computed by their respective Bayes factors, is only at $\sim 2.4\sigma$, preventing us from deriving further conclusions.

\section{Discussion}\label{sec:disc}

%------------------------------------------- LIGHT CURVE
    \begin{figure}
    \centering
    \includegraphics[width=9cm]{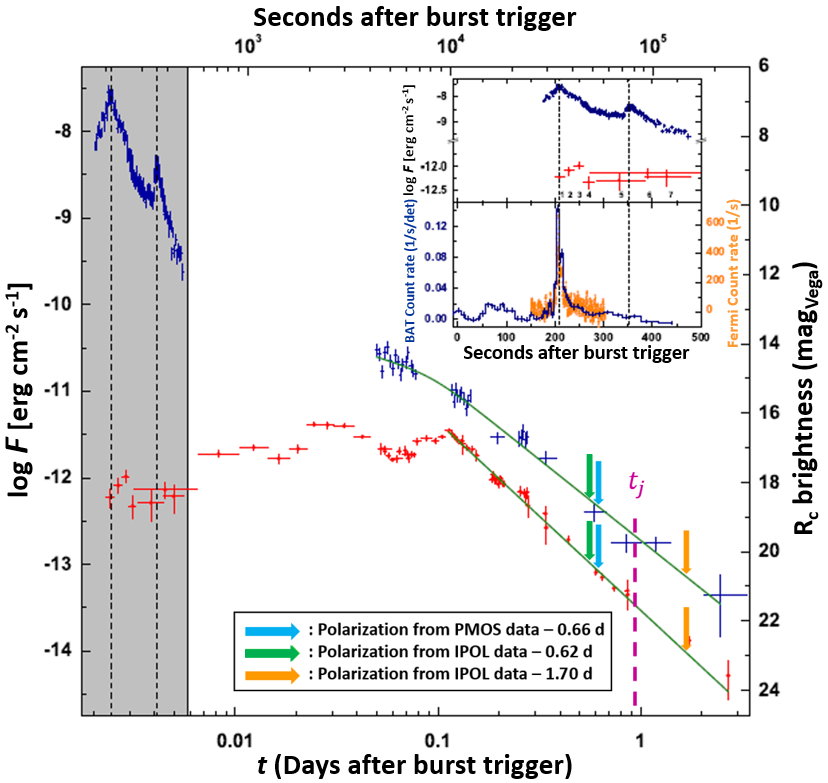}
    \caption{Optical (red points) and X-ray (blue bars, in 0.3-10 keV band) light curve of GRB\,080928 afterglow. The green curve is the best fit of the late-time data, while the grey-shaded region on the left shows the early phase. The latter is also represented in the inset and compared with the prompt-phase emission detected by Fermi-GBM and $\textit{Swift}$/BAT. The arrows above the best fit curves on the right represent the epochs of polarimetric observations obtained with ESO-VLT FORS1, both in the imaging polarimetry and spectro-polarimetry mode, and the assumed jet break time \textit{t$_j$} is indicated by the purple dashed line. From \cite{ROSSI_A11}.}
    \label{lc}
    \end{figure}
%-----------------------------------------------------------------

    In the previous section, we discuss  the possibility that the total polarisation we detect could be the combination of the intrinsically polarised radiation from the GRB optical afterglow and an additional contribution from dust in the host galaxy and/or in the Milky Way, which can be constrained by analysing the spectral dependence of the polarisation degree. The analysis of PMOS data allowed us to derive only an upper limit for $P$, possibly pointing to intrinsically unpolarised radiation and a negligible contribution of the dust aligned along the line of sight. In principle, this low value may also be due to the combination of a dust-induced contribution and non-zero intrinsic polarisation, which cancel out due to a phase shift of their position angle with respect to each other \citep[e.g.][]{Lazzati020813}. However, the Serkowski fit we performed on the polarisation spectrum did not prove solid evidence of a relevant dust contribution. For this reason, we decided to adopt in the following discussion the scenario with the lowest number of parameters, and we consider the dust-induced polarisation as negligible.

   The time dependence of the polarisation degree reported in Fig.\ref{p_t} can be compared with the expected behaviour depending on some features of the emitting region, the jet geometry in particular. \cite{ROSSI_E04} studied different jet configurations, with and without the possibility of lateral expansion, and derived the corresponding expected light curves and polarisation curves. An extensive range of values for $P$ can be obtained depending on some parameters, e.g. the observer viewing angle and the jet break time. If the observer viewing angle is lower than the jet aperture angle, we expect a relatively low polarisation degree, from a few \% to a maximum of $\sim 25\%$ for $\theta_{obs}=\theta_{jet}$. For off-axis observations the larger the observing angle, the later and the stronger the polarisation peak: even $ P\sim 50\%$ can be achieved in principle, but such a large polarisation degree has never been observed. In general, the time evolution of the curves strongly depends on the jet configuration, therefore it is important to perform multi-epoch observations to properly compare the observed polarisation curve with the theoretical expectations and possibly infer the intrinsic properties of the jet and the emitting region.
    
   These models are based on random magnetic fields that are confined to the shock, while \cite{Teboul&Shaviv21} also considered more complex models of the magnetic field configuration, from which they derived some polarisation predictions. In addition to the situation considered by \cite{ROSSI_E04}, \cite{Teboul&Shaviv21} computed expectations for an anisotropic random field, yielding a similar behaviour, that is two peaks separated by a 90$^{\circ}$ position angle swing for an observer inside the jet. Then, they added to the random field an additional, ordered component (still confined to the shock) whose relative importance is set by the random-to-ordered ratio, $\mu$. \cite{Teboul&Shaviv21} observed that the polarisation peak amplitude for off-axis observers is not significantly affected by $\mu$, while the value away from the maximum is much more sensitive to it. They finally added an anisotropy factor to the previous expectations, yielding a significant difference in the case where $\theta_{obs} > \theta_{jet}$, while for an observer inside the jet, the expectations do not depend to a significant extent  on the anisotropy factor. The rotation of the position angle is analysed as well: the 90$^{\circ}$ swing is expected for observers inside the jet axis and only if the field is totally random, whereas the rotation is smaller if we add an ordered component. Its value depends on the direction and the intensity of the ordered field.
    
    The light curve obtained for GRB\,080928 \citep[Fig.4 in][]{ROSSI_A11} is shown in Fig.\ref{lc}, both in X-rays and optical band. We identify a power-law decay. The X-ray data can be fit after $t$ = 4.2 ks from the detection with a broken power-law with indices $\alpha_1^X$ = 0.72 $\pm$ 0.35 and $\alpha_2^X$ = 1.87 $\pm$ 0.07, and a break time $t_b^X = 0.094 \pm 0.019$ d (all errors are 1$\sigma$ uncertainties). The optical curve follows a single power-law decay with index $\alpha^{opt}$ = 2.17 $\pm$ 0.02 after $t$ = 10 ks, while the complex behaviour observed before can be due to the prolonged activity of the central engine or to the presence of a precursor observed at $t_0-90$ s (see Sect. \hyperref[sec:GRB]{2}). If both the precursor and the main event originate from the same central engine, we could expect a non-standard afterglow with multiple peaks \citep[for more details, see][]{Nappoetal14}. However, we can observe this evolution in the optical data only, and the lack of X-ray data from $\sim$ 0.5 ks to 3 ks prevents us from discussing this scenario in more detail.
    
    The interpretation of the results and the comparison with expected models strongly depend on the jet break time $t_j$, which is expected to occur at different times depending on the viewing angle of the observer, the geometry of the jet, and the potential sideways expansion. According to the most popular afterglow models, the jet break time is achromatic, and therefore we should have $t_j^X = t_j^{opt}$, and we expect the X-ray and optical light curves to have the same decay index after the jet break. GRB\,080928 light curves are not consistent with this behaviour because $\alpha_2^X \neq \alpha^{opt}$. Moreover, we were not able to identify a break at $t_b^X$ in the optical curve due to its peculiar evolution. Even if a different index decay after the jet break is not an unprecedented behaviour \citep[see][]{Kannetal10, Zaninonietal13, Melandrietal14}, according to \citet{Leventisetal14} the break pointed out in the X-ray curve at $t_b^X$ is likely to be interpreted as an injection break. \citet{Leventisetal14} also assume the jet break to occur at $t_j = 80$ ks $ = 0.93$ d, which is in between our polarimetry epochs, as shown by the arrows and the dashed line in Fig.\ref{lc}.
    
    By comparing our light curve and polarisation curve with the expected ones from different jet geometries presented in \citet{ROSSI_E04}, we can identify the range of values expected for $P$. In the case of a homogeneous jet, we expect two maxima in the polarisation curve (the second always larger than the first), with a period of null polarisation and 90$^{\circ}$ rotation of the position angle in between. Depending on the parameter for the jet configuration, at some viewing angle this time should coincide with the jet break time, and slightly later or before at different $\theta_{obs}$. If the burst is observed at an angle $\theta_{obs}<\theta_{jet}$, the maxima of the curve can vary from $P\sim2\%$ to $P\sim20\%$, depending on the observer's viewing angle and the presence of a sideways expanding (SE) or a non-sideways expanding (NSE) jet. In particular, we see that the larger $\theta_{obs}/\theta_{jet}$, the earlier the jet break (and the period of null polarisation) is expected to occur. If $t_j = 0.93$ d is the actual jet break, we expect to have $P$ close to zero at $t \sim t_j$ for  $0.6\leq\theta_{obs}/\theta_{jet}\leq0.8$. The polarisation degree on the first night would be in the $\sim1\%-4\%$ range, while on the second night it would range from $\sim 4\%$ to a maximum value of $\sim 8\%$ for $\theta_{obs}/\theta_{jet} = 0.8$. These values are derived from the NSE-jet model, whereas for a SE jet the expected values are slightly smaller (by $\sim 1\%$). Alternatively, in the case of a structured or Gaussian jet, we expect similar non-zero polarisation before and after the jet break, which is at odds with our results. Only additional observations at earlier time might have allowed us a better comparison with the expectations from these jet configurations. A proper modelling is also difficult here because the polarisation curves depend on the opening angle of the jet core, a parameter that we cannot constrain for GRB\,080928. We can conclude that the expected values for $P$ for a homogeneous jet are consistent with our results: both the upper limits for the first night and the detection on the second night are within the low end of the range of expected values. The detection would be placed on the second rise of the polarisation curve. Therefore, we can state that we probably observed a homogeneous jet close to the jet axis, but we do not have enough data to discriminate between a NSE and a SE jet. Earlier observations would have provided additional data to look for a polarisation different from zero before $t_j$ (i.e. possibly coincident with the first peak) and a potential 90$^\circ$ swing of the position angle, as observed in GRB\,091018 \citep{Wiersemaetal12} and GRB\,121024 \citep{Wiersemaetal14}. Unfortunately, no data were taken before $t- t_0 \sim 0.62$ d.
    
    Comparison with models derived by \cite{Teboul&Shaviv21} cannot yield quantitative results because we would need a large number of parameters that we cannot constrain for this source, making it impossible to distinguish between any of the proposed configurations. We also compared our results with the sample of afterglows with both a jet break and a polarisation detection presented by \cite{StringerLazzati20}: GRB\,080928 shows a larger polarisation degree than the average, but it is still consistent with the sample within the uncertainties.
    
    Therefore, as mentioned above, our results are consistent with the scenario of a homogeneous jet, while only a larger dataset would have allowed us to make a comparison with the predictions from more complex jet structures. In addition to that, from the comparison with the expected polarisation curves computed by \cite{ROSSI_E04} and \cite{Teboul&Shaviv21}, we derived that the presence of significantly polarised radiation on the second night suggests that the emitting region was likely characterised by some degree of locally ordered magnetic fields. Indeed, we can generally identify locally ordered fields when we have a non-zero polarisation degree \citep{Granot&Konigl03}, while a null value is probably associated with a random magnetic field configuration, although our sparse dataset does not allow us to draw more quantitative conclusions.

\section{Conclusions}\label{sec:concl}

    We present multi-epoch observations and a polarisation study of the optical afterglow of GRB\,080928, which was first detected by $\textit{Swfit}$ and Fermi telescopes. The ground-based polarimetric follow-up performed by means of the ESO-VLT FORS1 instrument allowed us to perform a polarimetric analysis of the source in both imaging and spectro-polarimetry modes, a significant result in itself because, from a total of more than 1500 GRBs observed by $\textit{Swift}$ to date, IPOL data were only collected for about 20 of them, and PMOS data were collected for only four (those cited in Sect.\hyperref[sec:intro]{1}). This kind of measurement was made possible by the brightness of the event, which allowed us to collect data up to a couple of days after the trigger of the GRB. Thanks to polarimetry, we can investigate additional features of the emitting region and try to constrain, for example, the jet aperture angle and the local magnetic field configuration. 
    
    In the specific case of GRB\,080928, the modelling was not fully satisfactory; for example, the afterglow light curves do not follow the standard afterglow models. The analysis performed in this work led to the addition of another source in the limited sample of GRBs studied using the polarimetry technique: we have evidence of polarised radiation after $\sim$1.7 d from the initial detection at a 4$\sigma$ CL, with the polarisation curve rising towards this value from the non-polarised radiation detected on the first night. Detailed spectral modelling did not allow us to single out the possible contributions from the dust aligned along the line of sight and the intrinsic polarisation of the OA to the total polarisation degree. Indeed, the fit of the polarisation spectrum could be consistent with both a pure afterglow model and a Serkowski-like behaviour (see Sect. \ref{sec:res}). This result, together with the fact that we observe a significant variation for $P$, suggests that the observed polarisation is largely intrinsic. Moreover, if the jet break is fixed at $t_j = 0.93$ d, we may conclude that we observed a homogeneous jet slightly off-axis, with $0.6\leq\theta_{obs}/\theta_{jet}\leq0.8$. The lack of both earlier and later time measurements prevents us from deriving additional conclusions linked to the behaviour of the position angle and to the further evolution of both the light curve and the polarisation curve, which could in principle reveal more information about the source. Moreover, the availability of X-ray data between 0.5 ks and 3 ks would have allowed us to compare the observations with the theoretical X-ray curves expected from the high-latitude emission (HLE) from a structured jet seen off-axis. Indeed, by constraining the  viewing angle of the observer from polarisation measurements, we could in principle test the theoretical models developed to possibly explain the plateau often observed in the X-ray light curves of GRBs \citep[see][]{Ascenzietal20}. This could be an interesting comparison for the future detection of polarised radiation from GRBs.
    
    To date, polarisation measurements have been obtained with a rather irregular and, in some cases, limited temporal sampling. Complete coverage of both the light curve and the polarisation curve would permit a complete analysis. In particular, the opportunity to observe polarisation from very bright GRBs before and after the jet break from more bursts would allow us to compare them to the theoretical models in a very detailed way. Long-time polarisation observations with ground-based telescopes are a very demanding task, because they require a lot of observation time, but would be worthwhile given the progress that could be made in the field of GRB physics as a result. In particular, multi-epoch spectro-polarimetric observations could help to single out contributions from the host galaxy and the Milky Way, resulting in a refined measurement of the GRB intrinsic polarisation degree $P$ and could allow analysis of the polarisation properties of the dust in distant galaxies, a poorly studied feature in such systems. In addition, more precise measurements could help us to follow the evolution of both $P$ and $\theta$, which may help in putting more stringent constraints on the micro-physics parameters driving the synchrotron emission; on the jet geometry (e.g. the aperture angle, the Lorentz factor $\Gamma$); and on the specific configuration of local magnetic fields. Extending this kind of analysis to a large number of bursts would provide a wider sample with which to analyse and possibly realise a statistical study in the future.
    
    We show that with multi-epoch imaging polarimetry and spectro-polarimetry analyses, it is possible to derive the intrinsic polarisation of the optical afterglow at different times and wavelengths and to study the structure of GRB outflows, which will hopefully lead to a better understanding of the physical processes that give rise to these extremely bright astrophysical sources.
   
\begin{acknowledgements}
    The authors thank the referee for the very constructive comments and  Gianpiero Tagliaferri for useful discussions.
    The present work is based on observations collected at the European Southern Observatory under ESO programme 081.D-0192. SC, PDA and AM acknowledge funding from the Italian Space Agency, contract ASI/INAF n. I/004/11/5.
    KW acknowledges support through a UK Research and Innovation Future Leaders Fellowship (MR/T044136/1) awarded to dr.~B. Simmons.
\end{acknowledgements}

\bibliographystyle{aa}
\bibliography{biblio}

\begin{appendix}

\section{The dataset}
%-------------------- IPOL OBS. LOG
    \begin{table}[htp]
    \caption{Log of ESO-VLT FORS1 imaging polarimetry observations}
    \centering
    \begin{tabular}{cccc}
    \hline
    \noalign{\smallskip}
    TIME (UT)& t$_\text{exp}$[s]& ANGLE[$^{\circ}$]& FILTER\\
    \noalign{\smallskip}
    \hline
    \hline
    \noalign{\smallskip}
    \multicolumn{4}{c}{FIRST NIGHT - 09/29/08}\\
    \noalign{\smallskip}
    \hline
    \noalign{\smallskip}
    05:01:57& 330& 0& V\_HIGH\\
    05:08:00& 330& 22.5& V\_HIGH\\
    05:14:03& 330& 45&  V\_HIGH\\
    05:20:06& 330& 67.5& V\_HIGH\\
    05:26:32& 330& 0& V\_HIGH\\
    05:32:34& 330& 22.5& V\_HIGH\\
    05:38:38& 330& 45& V\_HIGH\\
    05:44:41& 330& 67.5& V\_HIGH\\
    \noalign{\smallskip}
    \hline
    \noalign{\smallskip}
    \multicolumn{4}{c}{SECOND NIGHT - 09/30/08}\\
    \noalign{\smallskip}
    \hline
    \noalign{\smallskip}
    06:25:05& 330& 0& V\_HIGH\\
    06:31:08& 330& 22.5& V\_HIGH\\
    06:37:11& 330& 45& V\_HIGH\\
    06:43:14& 330& 67.5& V\_HIGH\\
    06:49:39& 330& 0& V\_HIGH\\
    06:55:42& 330& 22.5& V\_HIGH\\
    07:01:45& 330& 45& V\_HIGH\\
    07:07:48& 330& 67.5& V\_HIGH\\
    \hline
    07:19:29& 330& 0& V\_HIGH\\
    07:25:32& 330& 22.5& V\_HIGH\\
    07:31:34& 330& 45& V\_HIGH\\
    07:37:37& 330& 67.5& V\_HIGH\\
    07:44:02& 330& 0& V\_HIGH\\
    07:50:05& 330& 22.5& V\_HIGH\\
    07:56:07& 330& 45& V\_HIGH\\
    08:02:10& 330& 67.5& V\_HIGH\\
    \hline
    08:14:00& 330& 0& V\_HIGH\\
    08:20:02& 330& 22.5& V\_HIGH\\
    08:26:05& 330& 45& V\_HIGH\\
    08:32:09& 330& 67.5& V\_HIGH\\
    08:38:33& 330& 0& V\_HIGH\\
    08:44:36& 330& 22.5& V\_HIGH\\
    08:50:38& 330& 45& V\_HIGH\\
    08:56:40& 330& 67.5& V\_HIGH\\
    \noalign{\smallskip}
    \hline
    \noalign{\smallskip}
    \end{tabular}
    \tablefoot{GRB\,080928 optical afterglow observations taken with the imaging polarimetry (IPOL) mode at ESO-VLT FORS1, on the nights of 2008 September 28 and 2008 September 29. There are a total of eight sets, two taken on the first night after the GRB event (from $\sim$ 14 h later) and six on the following night (from 39.39 h after the burst).}
    \label{data_ipol}
    \end{table}

%-------------------- PMOS OBS. LOG w/SEEING, AIR MASS
    \begin{table}[htp]
    \caption{Log of ESO-VLT FORS1 spectro-polarimetry observations}
    \centering
    \begin{tabular}{clccc}
    \hline
    \noalign{\smallskip}
    TIME& ANGLE& t$_\text{exp}$& SEEING& AIR MASS\\
    (UT-09/29)& [$^\circ$]& [s]& [$''$]& (start/end)\\
    \noalign{\smallskip}
    \hline
    \hline
    \noalign{\smallskip}
    05:58:22& 0& 300& 1.5& 1.85/1.81\\
    06:04:06& 22.5& 300& 1.5& 1.81/1.77\\
    06:09:51& 45& 300& 1.5& 1.77/1.74\\
    06:15:36& 67.5& 300& 1.5& 1.73/1.70\\
    06:21:43& 0& 300& 1.4& 1.67/1.67\\
    06:27:27& 22.5& 300& 1.5& 1.67/1.64\\
    06:33:11& 45& 300& 1.5& 1.63/1.61\\
    06:38:55& 67.5& 300& 1.5& 1.61/1.58\\
    06:51:49& 0& 300& 1.4& 1.55/1.52\\
    06:57:33& 22.5& 300& 1.5& 1.52/1.50\\
    07:03:18& 45& 300& 1.5& 1.50/1.48\\
    07:09:03& 67.5& 300& 1.5& 1.48/1.46\\
    07:15:11& 0& 300& 1.5& 1.45/1.44\\
    07:20:55& 22.5& 300& 1.4& 1.43/1.42\\
    07:26:39& 45& 300& 1.5& 1.41/1.40\\
    07:32:24& 67.5& 300& 1.2& 1.40/1.38\\
    \noalign{\smallskip}
    \hline
    \noalign{\smallskip}
    \end{tabular}
    \tablefoot{GRB\,080928 optical afterglow spectro-polarimetry (PMOS) observations taken with ESO-VLT FORS1 on the night of 2008 September 28, with seeing and air mass values. The results for the seeing lie within a 5\% error range with respect to a reference value of 1.5$''$, except for the last measurement.}
    \label{seeing}
    \end{table}

\end{appendix}

\end{document}